\documentclass[preprint,5p]{elsarticle}

\usepackage{graphicx}
\usepackage{amssymb,amsthm}

\journal{Rev. Bras. Ensino Fis.}

\begin{document}

\begin{frontmatter}

\title{Simulation of deterministic compartmental models for infectious diseases
dynamics}
  
\author{Antonio M. Batista$^{1,2,3}$, Silvio L. T. de Souza$^4$, Kelly C.
Iarosz$^{3,5,6,*}$, Alexandre C. L. Almeida$^7$, Jos\'e D. Szezech Jr$^{1,2}$,
Enrique C. Gabrick$^1$, Michele Mugnaine$^8$, Gefferson L. dos Santos$^{1,5}$,
Iber\^e L. Caldas$^3$}
\address{$^1$Postgraduate Program in Sciences, State University of Ponta Grossa,
84030-900, Ponta Grossa, PR, Brazil}
\address{$^2$Department of Mathematics and Statistics, State University of
  Ponta Grossa, 84030-900, Ponta Grossa, PR, Brazil}
\address{$^3$Physics Institute, University of S\~ao Paulo, 05508-090, S\~ao
  Paulo, SP, Brazil}
\address{$^4$Federal University of S\~ao Jo\~ao del-Rei, Campus Centro-Oeste,
  Divin\'opolis, MG, Brazil}
\address{$^5$Faculdade de Tel\^emaco Borba, FATEB, Tel\^emaco Borba, PR, Brazil}
\address{$^6$Graduate Program in Chemical Engineering Federal Technological
  University of Paran\'a, Ponta Grossa, PR, Brazil}
\address{$^7$Statistics, Physics and Mathematics Department, Federal University
  of S\~ao Jo\~ao del-Rei, Ouro Branco, MG, Brazil}
\address{$^8$Department of Physics, Federal University of Paran\'a, Curitiba,
PR, Brazil}

\cortext[cor]{kiarosz@gmail.com}

\date{\today}

\begin{abstract}
Infectious diseases are caused by pathogenic microorganisms and can spread
through different ways. Mathematical models and computational simulation have
been used extensively to investigate the transmission and spread of infectious
diseases. In other words, mathematical model simulation can be used to analyse
the dynamics of infectious diseases, aiming to understand the effects and how
to control the spread. In general, these models are based on compartments, where
each compartment contains individuals with the same characteristics, such as
susceptible, exposed, infected, and recovered. In this paper, we cast further
light on some classical epidemic models, reporting possible outcomes from
numerical simulation. Furthermore, we provide routines in a repository for
simulations.
\end{abstract}


\end{frontmatter}


\section{Introduction}

Infectious diseases have caused epidemics with devastating effects, for instance
influenza A virus sub-type H1N1 \cite{spreeuwenberg18} and smallpox
\cite{henderson11}. Infectious disease with pandemic potential is one of the
greatest challenges of the health system. Recently, the World Health
Organization declared COVID-19 \cite{chen20,han20} as a pandemic. The mortality
depends on many factors such as number of infected people, virulence, and
prevention efforts \cite{zhou20}. Various infectious diseases can come in waves,
e.g., the first three waves of avian influenza A (H7N9) virus circulation
\cite{xiang16}. Kissler et al. \cite{kissler20} projected that recurrent
outbreaks of COVID-19 will probably happen. 

Epidemiological models \cite{bailey57} have been proposed to analyse the spread
of infectious diseases in host populations \cite{fan01}. In 1760, Bernoulli
\cite{bernoulli60} proposed a model to describe the impact of variolation. In
1906, a mathematical model to explain the epidemic of measles was introduced by
Hamer \cite{hamer06}.

The SI model \cite{lopez07} describes the evolution of susceptible S and
infected I individuals, respectively. A model was proposed by Ross \cite{ross16}
in 1916 for malaria. In the Ross model, known as SIS model \cite{tome20b},
susceptible become infected and infected recover without immunity. Kermack and
McKendrick \cite{kermack27} introduced a model, known as SIR model, in which the
removed can be recovered, immune, or dead. The SIRS model \cite{souza10} was
obtained when a waning immunity was incorporated. In the SEIR model \cite{li01}
there are four states, where E corresponds to exposed, namely a latency period
is considered. It has been applied to measles \cite{grenfell94} and rubella
\cite{buonomo11}. The SEIRS model \cite{bjornstad20} considers individuals that
are transferred from the recovered to the susceptible compartments.

One of the different types of controls of epidemic is the vaccination. The
vaccination controls have as main objective to remove by immunity the population
from the susceptible state \cite{sen12}. Gao et al. \cite{gao08} demonstrated
that pulse vaccination can be an effective strategy for the elimination of
infectious diseases. Investigating the transmission of tuberculosis by means of
epidemiological models,  Liu et al. \cite{liu17} reported that mixed vaccination
gives a rapid control.

When there is no vaccine for the infectious disease, the control strategies
are based on quarantine and isolation, for instance the COVID-19 epidemics
\cite{tang20}. Law et al. \cite{law20} studied a time-varying SIR model for the
transmission dynamics of COVID-19. Prem et al. \cite{prem20} reported that the
magnitude of the epidemic peak can be reduced through sustained physical
distancing. Feng \cite{feng07} analysed the final and peak epidemic sizes
considering quarantine and isolation in SEIR models. Recently, Boldog et al.
\cite{boldog20} analysed the risk assessment of novel coronavirus outbreaks
outside China by means of an extension of a standard SEIR model. 

Stochastic mathematical models have been considered in prediction of the spread
of several infectious diseases \cite{tome20a}. The law of mass-action has been
used for almost a century in epidemiology to describe the contact rate of
individuals. This law originated in practice and theory of chemical reaction
kinetics \cite{tome15}. Tom\'e and Oliveira \cite{tome20b} presented models
for epidemic spreading and showed the analogy between the spreading of a
disease with a critical phase transition. They also analysed the epidemic curve,
which is a graphical representation of the number of identified cases over a
period of time.

In this work, we focus on the simulation of different deterministic models that
have been used to describe infectious disease dynamics. We provide routines in
a repository \cite{repository} to assist students with their first steps about
computer simulations of mathematical models in epidemiology. The reader can find
mathematical details in Ref. \cite{tome20b}. We show the dynamical behaviours
predict by some mathematical models, such as SI, SIS, SIR, SIRS, SEIR, and
SEIRS. We present an application of the SIR model in COVID-19 for the parameters
obtained from China according to Ref. \cite{cooper20}. Considering the SEIR
model, we show the impact of easing restriction on the infection rate that was
reported in Ref. \cite{souza21}.


\section{The SI model}

\begin{figure}[htb]
\centerline{\includegraphics[width=0.4\linewidth]{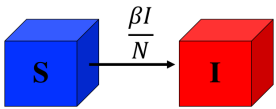}}
\caption{Compartment diagram of the SI model, where S and I are the numbers of
susceptible and infected individuals, respectively, and $\beta$ (in units of
1/day) is the effective contact rate of the disease.}
\label{fig1}
\end{figure}

The SI model describes the dynamical behaviour of transmitted diseases
\cite{barros03} with interactions among infected and susceptible people. In this
model, vital processes are not considered, such as rates of birth and mortality.
Figure \ref{fig1} displays the diagram of the SI model, where the population is
divided into susceptible and infected individuals. After infected, the
individual does not return to the susceptible class, for instance herpes that
is spread from person to person by the virus Herpesviridae.

\begin{figure}[htb]
\centerline{\includegraphics[width=1\linewidth]{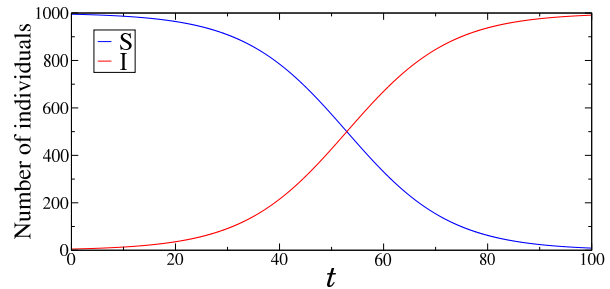}}
\caption{Time evolution of $S$ (blue line) and $I$ (red line) for
$N=1000$, $S(0)=995$, $I(0)=5$, and $\beta=0.1$.}
\label{fig2}
\end{figure}

The SI model is given by
\begin{eqnarray}
\frac{dS}{dt} & = & -\frac{\beta SI}{N}, \\
\frac{dI}{dt} & = & \frac{\beta SI}{N},  
\end{eqnarray}
where $N=S+I$ is the total population and $\beta$ is the effective contact rate
of the disease. Figure \ref{fig2} exhibits the time evolution (in days) of $S$
(blue line) and $I$ (red line) for $\beta=0.1$. With time, every susceptible
individual becomes infected.

\section{The SIS model}

\begin{figure}[htb]
\centerline{\includegraphics[width=0.4\linewidth]{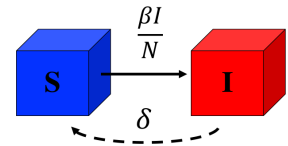}}
\caption{Compartment diagram of the SIS model, where S and I are the numbers of
susceptible and infected individuals, respectively, $\beta$ (in units of 1/day)
is the infectious rate, and $\delta$ (in units of 1/day) is the rate in which
the infected individuals recover to the susceptible state.}
\label{fig3}
\end{figure}

In the SIS model \cite{hethcote95,nakamura19}, which was studied recently
\cite{tome20b,tome20a}, the infected individuals can become susceptible,
however, there is no long-lasting immunity. Due to this fact, an individual can
have recurrent infections, such as common cold (rhinoviruses) and influenza, as
well as sexually transmitted diseases, for instance chlamydia and gonorrhoea. In
Fig. \ref{fig3}, we see a schematic representation of the compartment diagram
related to the SIS model.

\begin{figure}[htb]
\centerline{\includegraphics[width=1\linewidth]{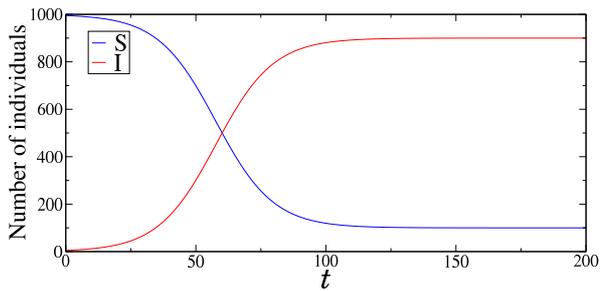}}
\caption{Time evolution of $S$ (blue line) and $I$ (red line) for $N=1000$,
$S(0)=995$, $I(0)=5$, $\beta=0.1$, and $\delta=0.01$.}
\label{fig4}
\end{figure}

The SIS model is written as
\begin{eqnarray}
\frac{dS}{dt} & = & -\frac{\beta SI}{N}+\delta I, \\
\frac{dI}{dt} & = & \frac{\beta SI}{N}-\delta I,  
\end{eqnarray}
where the infected individual goes to the susceptible state with a rate
$\delta$. In Fig. \ref{fig4}, we consider $\beta=0.1$ and $\delta=0.01$ for
$N=1000$. The SIS model has two stable equilibria, one for $I=0$ and another for
$I=N(1-\delta/\beta)$.

\section{The SIR model}

\begin{figure}[htb]
\centerline{\includegraphics[width=0.6\linewidth]{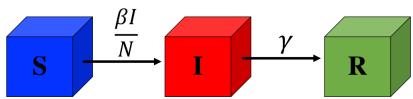}}
\caption{Compartment diagram of the SIR model, where S, I, and R are the numbers
of susceptible, infected, and recovered individuals, respectively, $\beta$ (in
units of 1/day) is the infectious rate, and $\gamma$ (in units of 1/day) is the
rate in which the infected individuals recover.}
\label{fig5}
\end{figure}

In 1927, Kermack and McKendrick \cite{kermack27} proposed the SIR model, a
mathematical model of spread of an infectious disease within a population,
studied recently \cite{tome20b,tome20a}. They considered three compartments, in
which the susceptible individuals go to the infectious compartment according to
an infectious rate and, depending on the recover rate, the infected individual
recover and develop immunity, as shown in Fig. \ref{fig5}.

The SIR model was introduced to explain the rapid increase of infected
individuals, as verified in epidemics such as the cholera and the plague. The
mathematical model is given by
\begin{eqnarray}
\frac{dS}{dt} & = & -\frac{\beta SI}{N}, \\
\frac{dI}{dt} & = & \frac{\beta SI}{N}-\gamma I, \\
\frac{dR}{dt} & = & \gamma I, 
\end{eqnarray}
where $\gamma$ is the recovery rate and $N=S+I+R$. The behaviour of the
infectious class has a dependence on the parameter $R_0=\beta/\gamma$, known as
reproduction ratio. $R_0$ is an important threshold quantity that describes the
transmissibility or contagiousness of pathogenic microorganisms
\cite{heffernan05}. In a susceptible group of individuals, it gives the expected
number of secondary cases of infections due to an infected individual. The
pathogenic microorganism is able to invade the population of susceptible
individuals if $R_0>1$. The births and deaths are not considered due to the
fact that the infection and recovery rates are fast.

\begin{figure}[htb]
\centerline{\includegraphics[width=1\linewidth]{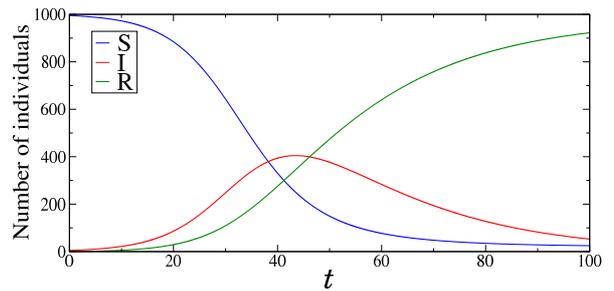}}
\caption{Time evolution of $S$ (blue line), $I$ (red line), and $R$ (green line)
for $N=1000$, $S(0)=995$, $I(0)=5$, $\beta=0.2$, and $\gamma=0.05$.}
\label{fig6}
\end{figure}

Figure \ref{fig6} shows the time evolution (in days) of $S$ (blue line), $I$
(red line), and $R$ (green line) for $N=1000$, $\beta=0.2$, and $\gamma=0.05$.
As initial conditions, we consider $S(0)=995$ and $I(0)=5$. We see that the
infected population reaches a peak while the susceptible individuals decrease
and recovered ones increase. After the peak, the number of infected individuals
decreases.

\begin{figure}[htb]
\centerline{\includegraphics[width=1\linewidth]{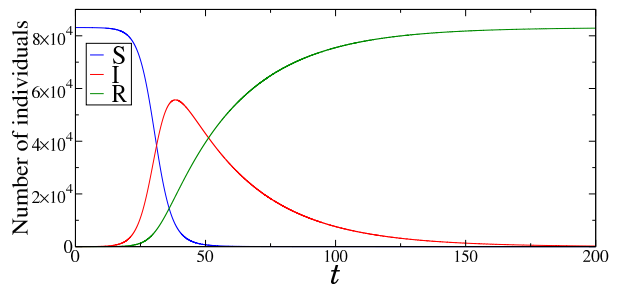}}
\caption{Time evolution of $S$ (blue line), $I$ (red line), and $R$ (green line)
for $N=83132$, $S(0)=83127$, $I(0)=5$, $\beta=0.35$, and $\gamma=0.035$.}
\label{fig7}
\end{figure}

In 2020, the SIR model was utilised by Cooper et al. \cite{cooper20} to study
the spread of COVID-19 in different communities. They analysed data recorded
between January and June 2020 from China, South Korea, India, Australia, Italy,
and the state of Texas in the USA. It was demonstrated that the SIR model can
provide a theoretical fra\-mework to study how the COVID-19 virus spreads within
communities. Figure \ref{fig7} displays the time evolution of $S$ (blue line),
$I$ (red line), and $R$ (green line) for $N=83132$, $S(0)=83127$, $I(0)=5$,
$\beta=0.35$, and $\gamma=0.035$, where the parameters are selected according to
Ref. \cite{cooper20} for China.

\section{The SIRS model}

\begin{figure}[htb]
\centerline{\includegraphics[width=0.6\linewidth]{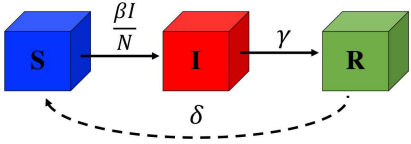}}
\caption{Compartment diagram of the SIRS model, where S, I, and R are the
numbers of susceptible, infected, and recovered individuals, respectively,
$\beta$ (in units of 1/day) is the infectious rate, $\gamma$ (in units of 1/day)
is the rate in which the infected individuals recover, $\delta$ (in units of
1/day) is the rate in which the recovered individuals return to the susceptible
class.}
\label{fig8}
\end{figure}

The SIRS model has been studied by various authors \cite{li17,song18} and
studied recently \cite{tome20b,souza10,tome20a}. Figure \ref{fig8} exhibits the
process diagram for the SIRS model. In this model, the recovery can generate
temporary immunity, and as a consequence the recovered individuals return to the
susceptible class after some time. The loss of immunity is observed in smallpox,
tetanus, and influenza.

The SIRS model is given by
\begin{eqnarray}
\frac{dS}{dt} & = & -\frac{\beta SI}{N}+\delta R, \\
\frac{dI}{dt} & = & \frac{\beta SI}{N}-\gamma I, \\
\frac{dR}{dt} & = & \gamma I-\delta R, 
\end{eqnarray}
where $\delta$ is the rate in which the recovered individuals return to the
susceptible class after losing the immunity and $N=S+I+R$.

\begin{figure}[htb]
\centerline{\includegraphics[width=1\linewidth]{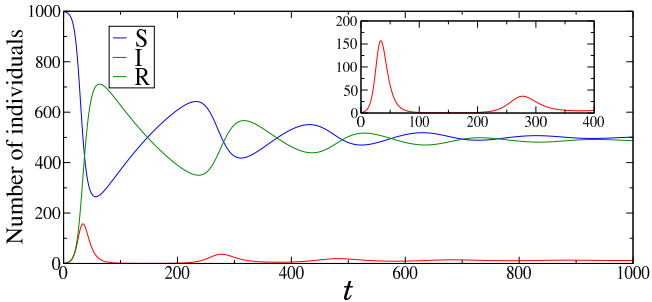}}
\caption{Time evolution of $S$ (blue line), $I$ (red line), and $R$ (green line)
for $N=1000$, $S(0)=995$, $I(0)=5$, $\beta=0.4$, $\gamma=0.2$, and
$\delta=0.005$. The inset figure corresponds to the magnification of the
infected individuals.}
\label{fig9}
\end{figure}

In Fig. \ref{fig9}, the blue, red, and green lines correspond to the time
evolution (in days) of $S$, $I$, and $R$, respectively. We see the persistence
of the infected population due to the transfer of individuals from the recovery
class to the susceptible one.

\section{The SEIR model}

\begin{figure}[htb]
\centerline{\includegraphics[width=0.9\linewidth]{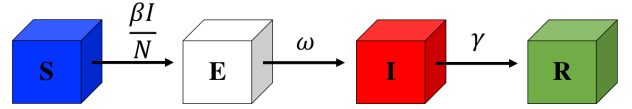}}
\caption{Compartment diagram of the SEIR model, where S, E, I, and R are the
numbers of susceptible, exposed, infected, and removed individuals,
respectively, $\beta$ (in units of 1/day) is the infectious rate, $\omega$ (in
units of 1/day) is the coefficient of migration rate, and $\gamma$ (in units of
1/day) is the rate in which the infected individuals recover.}
\label{fig10}
\end{figure}

One of the compartmental models of infectious diseases is the SEIR model,
studied recently \cite{tome20b,tome20a}. In this model, the individuals are
separated into four compartments. The SEIR model assumes people in the
susceptible (S), exposed (E), infected (I), and recovered (R) states, as shown
in Fig. \ref{fig10}.

The people move from S to E due to direct or indirect contact. In the E stage,
the people are infected but are not infectious, namely an latent period. The
infected individuals are recovered. The SEIR model was used by Carcione et al.
\cite{cardione20} to simulate the COVID-19 epidemic. 

The SEIR model is written as
\begin{eqnarray}
\frac{dS}{dt} & = & -\frac{\beta SI}{N}, \\
\frac{dE}{dt} & = & \frac{\beta SI}{N}-\omega E, \\
\frac{dI}{dt} & = & \omega E-\gamma I, \\
\frac{dR}{dt} & = & \gamma I,
\end{eqnarray}
where $\beta$ is the coefficient of infection rate, $\omega$ is the coefficient
of migration rate of latency, $\gamma$ is the coefficient of migration rate, and
$N=S+E+I+R$ is the total population.

\begin{figure}[htb]
\centerline{\includegraphics[width=1\linewidth]{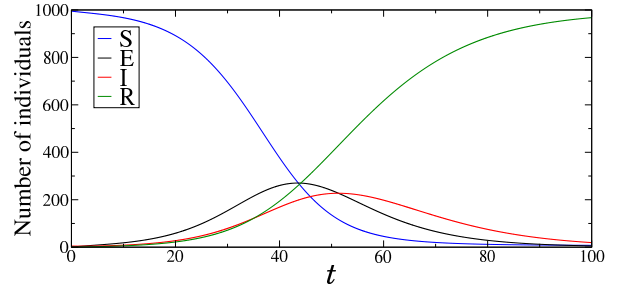}}
\caption{Time evolution of $S$ (blue line), $E$ (black line), $I$ (red line),
and $R$ (green line) for $N=1000$, $S(0)=995$, $I(0)=5$, $\beta=0.5$,
$\omega=0.1$, and $\gamma=0.1$.}
\label{fig11}
\end{figure}

Figure \ref{fig11} displays the temporal evolution (in days) of $S$ (blue line),
$E$ (black line), $I$ (red line), and $R$ (green line) for $\beta=0.5$,
$\gamma=0.01$, $\omega=0.1$. Initially, $S$ decreases and $E$, $I$, and $R$
increase. $S$ is depleted by the epidemic and goes to zero. Both $E$ and $I$
exhibit peaks and $R$ saturates when they go to zero.

\begin{figure}[htb]
\centerline{\includegraphics[width=0.65\linewidth]{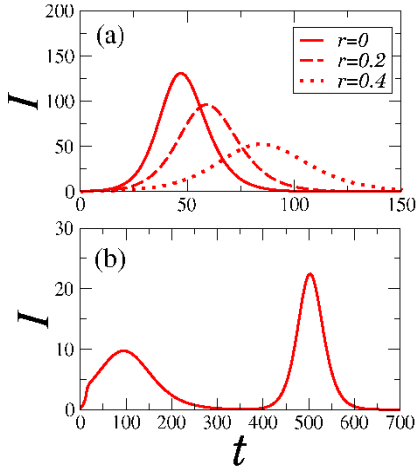}}
\caption{(a) Time evolution of $I$ for $\omega=0.2$, $\gamma=0.25$, $r=0$
(red line), $r=0.2$ (red dashed line), and $r=0.4$ (red dotted line). (b) Time
evolution of $I$ for $r=0$ ($t<14$), $r=0.6$ ($14\leq t<365$), and $r=0.2$
($t\geq 365$).}  
\label{fig12}
\end{figure}

Recently, Souza et al. \cite{souza21} considered the SEIR model to analyse the
impact of easing restriction on the infection rate during COVID-19 pandemic.
They included a parameter related to the restriction $r$ in the SEIR model,
$\beta\rightarrow \beta(1-r)$. By increasing $r$, the peak of infectious is 
delayed and the curve peak is flattened, as shown in Fig. \ref{fig12}(a). Figure
\ref{fig12}(b) shows that changes in the value of $r$ can generate a second
wave, namely the increase of the number of infected individuals after few cases.

\section{The SEIRS model}

\begin{figure}[htb]
\centerline{\includegraphics[width=0.9\linewidth]{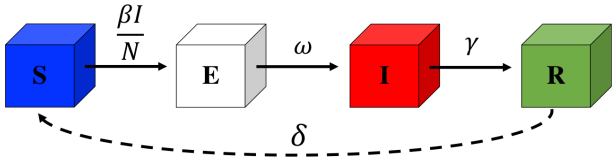}}
\caption{Compartment diagram of the SEIRS model, where S, E, I, and R are the
numbers of susceptible, exposed, infected, and removed individuals,
respectively, $\beta$ (in units of 1/day) is the infectious rate, $\omega$ (in
units of 1/day) is the coefficient of migration rate, $\gamma$ (in units of
1/day) is the rate in which the infected individuals recover, and $\delta$ (in
units of 1/day) is the rate in which the recovered individuals return to the
susceptible class.}
\label{fig13}
\end{figure}

In the SEIRS model, the susceptible individuals first go through the exposed
class before infected one. After the infectious, they are transferred to the
recovered compartment and become susceptible again, as shown in Fig.
\ref{fig13}. Denphedtnong et al. \cite{denphedtong13} used the SEIRS epidemic
model to describe the spread of diseases by transports. They investigated the
data of SARS (severe acute respiratory syndrome) outbreak in 2003.

The SEIRS model is given by
\begin{eqnarray}
\frac{dS}{dt} & = & -\frac{\beta SI}{N}+\delta R, \\
\frac{dE}{dt} & = & \frac{\beta SI}{N}-\omega E, \\
\frac{dI}{dt} & = & \omega E-\gamma I, \\
\frac{dR}{dt} & = & \gamma I-\delta R,
\end{eqnarray}
where $\delta$ is the rate in which the recovered individuals return to the
susceptible class after losing the immunity.

\begin{figure}[htb]
\centerline{\includegraphics[width=1\linewidth]{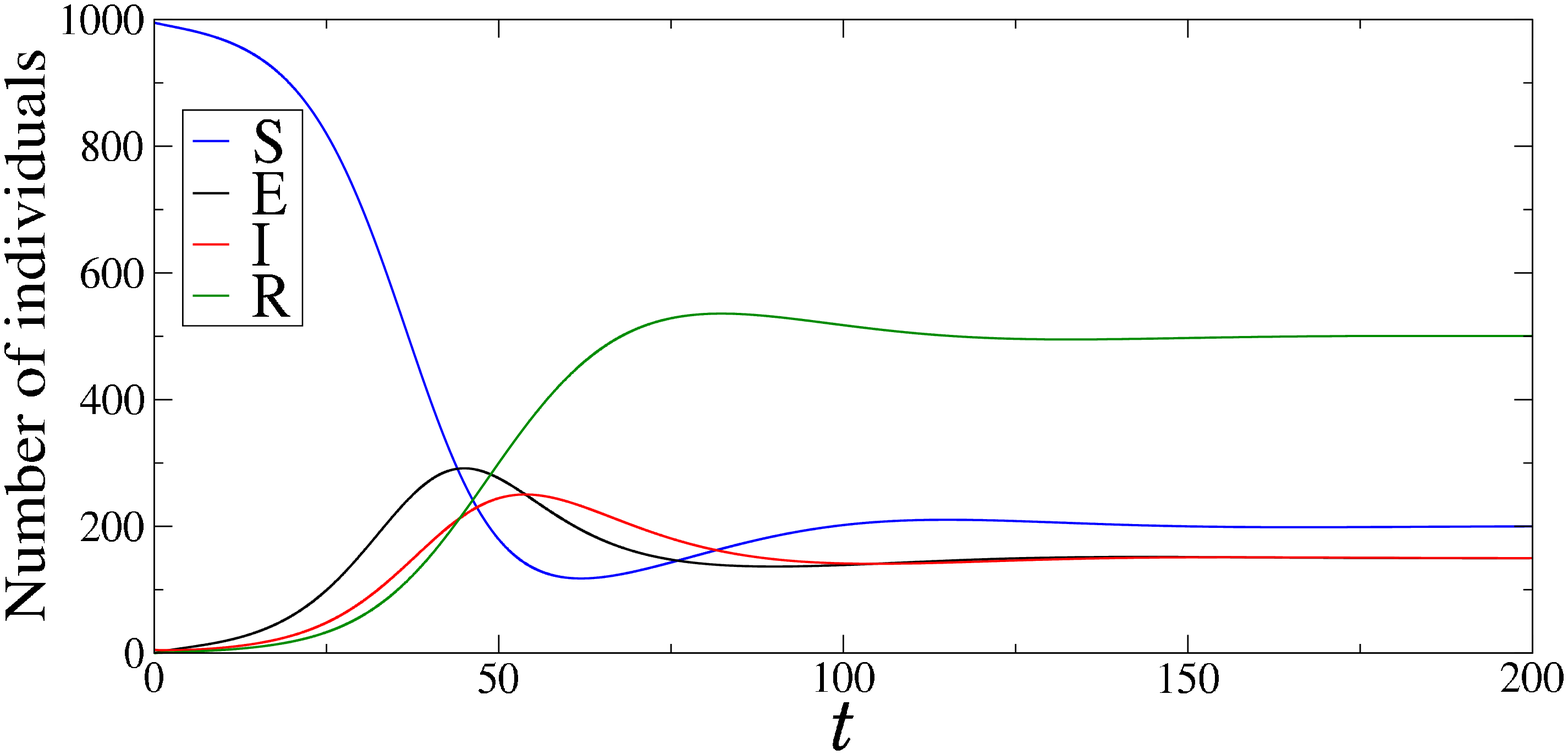}}
\caption{Time evolution of $S$ (blue line), $E$ (black line), $I$ (red line),
and $R$ (green line) for $N=1000$, $S(0)=995$, $I(0)=5$, $\beta=0.5$,
$\omega=0.1$, $\gamma=0.1$, and $\delta=0.03$.}
\label{fig14}
\end{figure}

In Fig. \ref{fig14}, we calculate $S$ (blue line), $E$ (black line), $I$ (red
line), and $R$ (green line) for $N=1000$, $S(0)=995$, $I(0)=5$, $\beta=0.5$,
$\omega=0.1$, $\gamma=0.1$, and $\delta=0.03$. We see that the susceptible,
exposed, and infected individuals do not go to the value equal to zero over
time. This occurs due to the fact that recovered individuals become susceptible
again. 


\section{Conclusions}

Investigations about infectious diseases play a crucial role in reducing
negative consequences and improving the recovery of individuals. Scientists have
been carried out research in epidemiology to understand how the people are
affected by transmissible diseases over time. Many different mathematical models
were proposed to find epidemiological parameters and manners to prevent illness.

In this work, we present some mathematical models that have been used to
describe the dynamical behaviour of infectious diseases. The models are based on
individuals that are separated into compartments, such as susceptible (S),
exposed (E), infected (I), and recovered (R). They focus on the prediction of
the population growth or reduction in each compartment. The models depend on the
parameters related to the rate in which individuals are transferred between the
classes.

The SI model considers that the individuals go from the susceptible to infected
states, while the SIS model assumes that there is no immunity and the infected
individuals return to the susceptible compartment. In the SIR and SIRS model,
it is included the class of the recovered individuals. The SEIR and SEIRS models
have not only the susceptible, infected, and recovered individuals, but also the
exposed ones. Recently, the SIR and SEIR models have been used in studies about
the COVID-19 epidemic. The SIR, SIRS, SEIR, and SEIRS models exhibit a peak of
infected people that was observed in different infection spread.

It has been demonstrated that through the intensity of restrictions associated
with the control policies to reduce the infection spread, it is possible to
flat the curve associated with the temporal evolution of the infected people
\cite{souza21}. However, the results depend on the duration and specific time
in which the restriction is applied. The peak of infections can be reduced by
means of restrictions or another peak can appear after suspending the
restrictions.

We provide routines in a repository for the SI, SIS, SIR, SIRS, SEIR, and SEIRS
models \cite{repository}.


\section*{Acknowledgements}
This work was possible by partial financial support from the following Brazilian
government agencies: CAPES, CNPq ($407543/2018-0$, $302903/2018-6$,
$420699/2018-0$, $407299/2018-1$, $428388/2018-3$, $311168/2020-5$),
Funda\c c\~ao Arauc\'aria, and S\~ao Paulo Research Foundation (FAPESP
$2018/03211-6$). We would like to thank www. 105groupscience.com.


\end{document}